\documentclass{PoS}

\title{KLOE-2 experiment at DA$\Phi$NE upgraded in luminosity}

\ShortTitle{KLOE-2 at upgraded DA$\Phi$NE}

\author{\speaker{Wojciech Wislicki}, on behalf of the KLOE-2 Collaboration\thanks{F. Archilli, D. Babusci, D. Badoni, G. Bencivenni, C. Bini, C. Bloise, V. Bocci, F. Bossi, P. Branchini, A. Budano, S. A. Bulychjev, P. Campana, G. Capon, F. Ceradini, P. Ciambrone, E. Czerwinski, E. Dane, E. De Lucia, G. De Robertis, A. De Santis, G. De Zorzi, A. Di Domenico, C. Di Donato, B. Di Micco, D. Domenici, O. Erriquez, G. Felici, S. Fiore, P. Franzini, P. Gauzzi, S. Giovannella, F. Gonnella, E. Graziani, F. Happacher, B. Hoistad, E. Iarocci, M. Jacewicz, T. Johansson, V. Kulikov, A. Kupsc, J. Lee-Franzini, F. Loddo, M. Martemianov, M. Martini, M. Matsyuk, R. Messi, S. Miscetti, D. Moricciani, G. Morello, P. Moskal, F. Nguyen, A. Passeri, V. Patera, A. Ranieri, P. Santangelo, I. Sarra, M. Schioppa, B. Sciascia, A. Sciubba, M. Silarski, S. Stucci, C. Taccini, L. Tortora, G. Venanzoni, R. Versaci, W. Wislicki, M. Wolke, J. Zdebik}\\
        A. Soltan Institute for Nuclear Studies, Laboratory for High Energy Physics,\\ Hoza 69, PL-00-681 Warszawa\\
        E-mail: \email{wislicki@fuw.edu.pl}}

\abstract{Prospective presentation is given for the experimental program of the KLOE-2 Collaboration, to be performed using the DA$\Phi$NE $e^+e^-$ collider upgraded in luminosity.
Data with the total luminosity of 25 fb$^{-1}$ are aimed to be collected in 3 years.
Major modifications of the accelerator and the spectrometer are described.
The KLOE-2 physics program contains: CKM unitarity and lepton universality tests, $\gamma\gamma$ physics, search for quantum decoherence and testing CPT conservation, low-energy QCD, rare kaon decays, physics of $\eta$ and $\eta^\prime$, structure of low-mass scalars, contribution of vacuum polarization to $(g-2)_{\mu}$, possible search for WIMP dark matter.
In this paper only selected physics subjects are reported.
}

\FullConference{The Xth Nicola Cabibbo International Conference on Heavy Quarks and Leptons,\\
		October 11-15, 2010\\
		Frascati (Rome) Italy}

\begin{document}

\section{Introduction}
\begin{figure}[h]
\begin{tabular}{ccc}
\includegraphics[scale=.19]{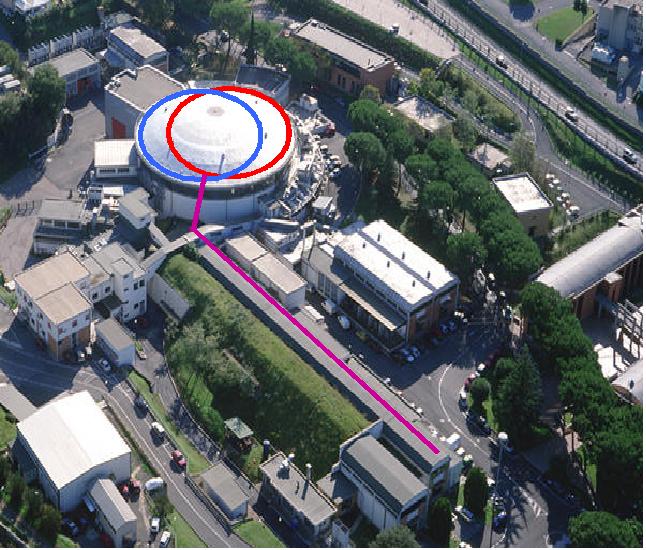} & \includegraphics[scale=.18]{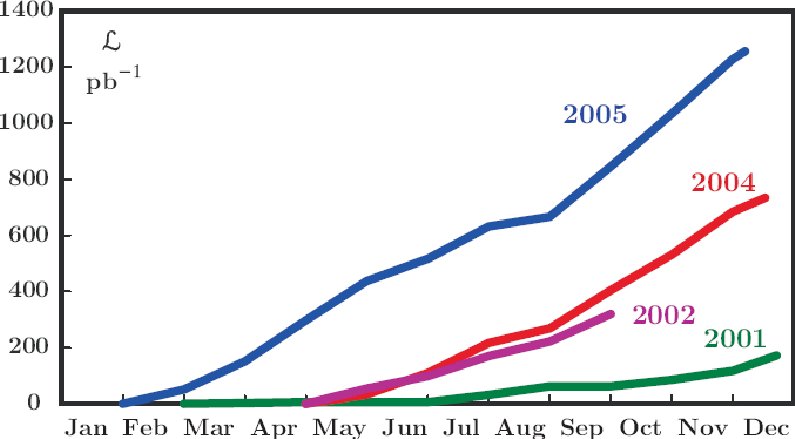} & \includegraphics[scale=.19]{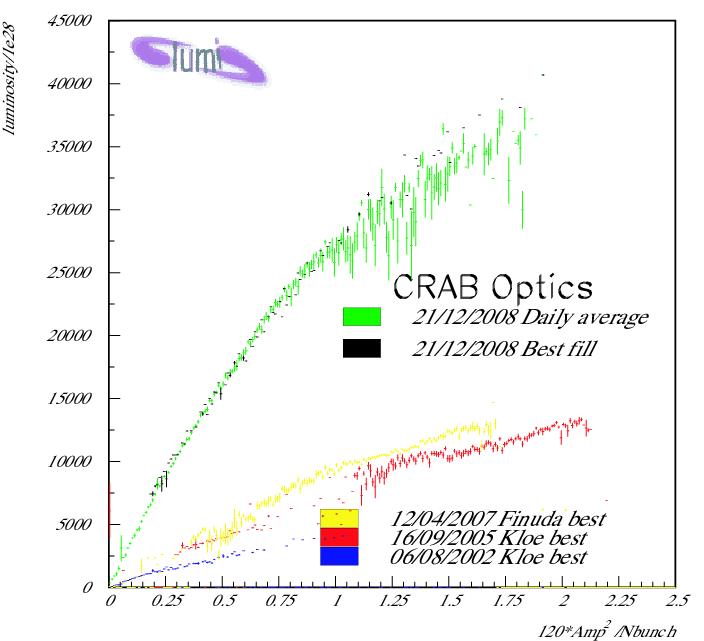}
\end{tabular}
\caption{The birds-eye view of the DA$\Phi$NE collider located at the INFN's LNF, where the $e^+$ and $e^-$ rings are schematically indicated (left), the cummulative year-by-year luminosities in KLOE experiment (centre), and the KLOE-2 luminosity with improved, crab waist, beam optics (right).}
\label{fig1}
\end{figure}

In 2005, KLOE Collaboration \cite{bossi1} has finished taking data with a total luminosity of 2.5 fb$^{-1}$.
The experiment was performed at the DA$\Phi$NE $e^+e^-$ collider (cf. Fig.~\ref{fig1}, left), owned by the INFN's Laboratori Nazionali di Frascati, and operating at the centre-of-mass energy equal to the $\Phi(1019)$ mass.
Fig.~\ref{fig1} (centre) presents year-by-year luminosities during the 5-year KLOE run.
Physics program approved for KLOE-2 \cite{beck1, amelino1}, the successor of KLOE, requires an order of magnitude increase of the volume of data and significant improvements in data quality. 
In order to achieve this, major upgrade of DA$\Phi$NE, ensuring its higher luminosity, was performed. 
The original KLOE detector is complemented with new detectors, enlarging its acceptance and improving accuracy.  
KLOE-2 begins taking data at the turn of 2010 and 2011, and aims to collect 25 fb$^{-1}$ in 3 years.

\section{DA$\Phi$NE luminosity and KLOE-2 detector upgrades}

Improvement of DA$\Phi$NE luminosity is realized by increasing Piwinski angle and implementing new beam optics \cite{zobow}.
For beams of the lateral size $\sigma_x$ at their crossing point and colliding at angle $\theta$, the Piwinski angle depends on them like $\varphi\sim\theta/\sigma_x$.
In order to make $\varphi$ larger, and thus shrinking beam overlap region, one increases $\theta$ and reduces $\sigma_x$.
Necessary compensation of betatron resonances is done by using two sextupoles at both sides of the interaction point and is called the {\it crab waist} optics.
Increase of $\theta$ by factor of 2 and $\sigma_x$ reduction by factor 2-3 effects with 3.4-4 times amplification of luminosity compared to 2005, as seen in Fig.~\ref{fig1} (right).

\begin{figure}[h]
\begin{tabular}{cc}
\hspace{1cm} \includegraphics[height=4cm,width=5cm]{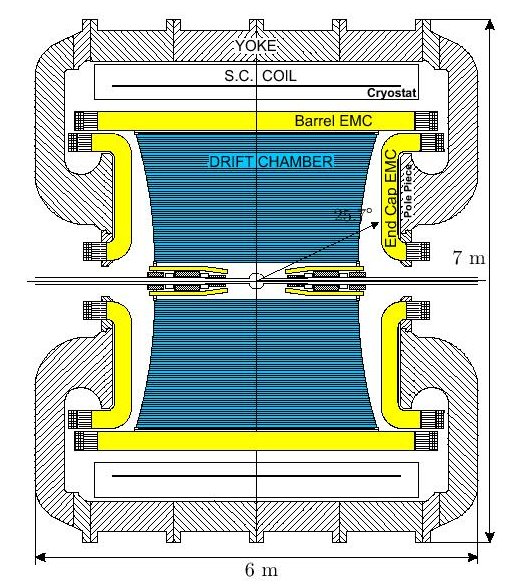} & \hspace{1cm} \includegraphics[height=4cm,width=5cm]{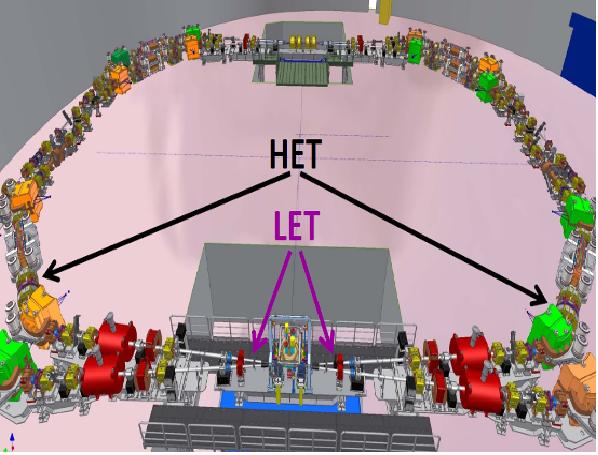} \\
\hspace{1cm} \includegraphics[height=4cm,width=5cm]{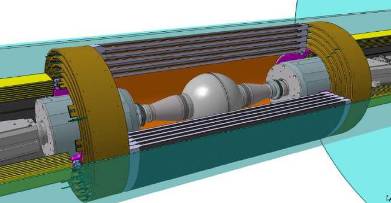} & \hspace{1cm} \includegraphics[height=4cm,width=5cm]{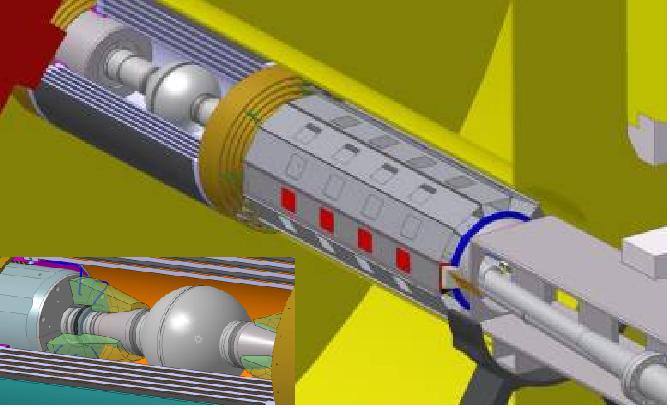}
\end{tabular}
\caption{Upper left: the KLOE spectrometer surrounding the $e^+e^-$ interaction point, consisting of drift chamber (blue), the barrel and endcap segments of the electromagnetic calorimeter (yellow), superconducting magnet coil embedded in cryostat and the magnet yoke.
Upper right: KLOE-2 electron taggers suitable for detection of lower-energy electrons of 160-230 MeV (LET) and higher-energy electrons of more than 400 MeV (HET).
Lower left: KLOE-2 inner tracking detector located close to the interacting point and consisting of four GEM layers.
Lower right: KLOE-2 near beam detectors: CCAL, for extending photon angular acceptance from 21$\,^\circ$ down to 10$\,^\circ$ (insert), and QCALT consisting of silicon photomultipliers, wave shifters and quadrupole magnets.}
\label{fig2}
\end{figure}

The KLOE spectrometer, covering 98\% of $4\pi$ solid angle, is shown in Fig.~\ref{fig2} (upper left).
The 0.52 T magnetic field and large drift chamber, filled 9:1 with helium and isobutane, ensure momentum resolution $\sigma(p_T)/p_T<0.4\,\%$ and 1 mm vertex resolution.
Sampling electromagnetic calorimeter, made of sandwitched lead and scintillating fibres, provides with $5.7\,\%/\sqrt{E\mbox{ (MeV)}}$ energy resolution and $(55/\sqrt{E}\oplus 100)\mbox{ ps}$ time resolution.

Additional equipments are installed by the KLOE-2, in order to meet requirements of physics measurements.
Research of hadronic final states produced in collisions of two photons with low virtuality, $e^+e^-\rightarrow e^+e^-\gamma^\ast\gamma^\ast\rightarrow e^+e^-X$, needs selective tagging of scattered electrons with momenta close to the beam momentum.
To this end, two pairs of taggers are installed: the high-energy tagger (HET), made of scintillators and photomultiplier tubes, designed for $E>400$ MeV, and the low-energy one (LET), made of inner LYSO crystals and silicon photomultipliers, capable of detection of $160<E<230$ MeV electrons (cf. Fig.~\ref{fig2}, upper right).
Improvement in vertex resolution and acceptance for particles with low $p_T$ is obtained by the inner tracking detector, made of four cyllindrical GEM layers around the beam pipe, as shown in Fig.~\ref{fig2} (lower left).
It provides with the 100 $\mu$m spatial resolution and a few ns time resolution, typically improving accuracy of decay vertex reconstruction by factor of 4.
New detectors are installed close to the beam line in order to enhance acceptance for photons at low angle and provide with additional improvement of reconstruction of decays far from the interaction point (Fig.~\ref{fig2}, lower right).
These are the CCAL, consisting of LYSO scintillating crystals with avalanche photodiodes, improving acceptance for photons from 21$\,^\circ$ to 10$\,^\circ$, and QCALT, made of silicon photomultipliers with wavelength shifters and additional quadrupoles.

\section{Selected issues of the KLOE-2 physics program}

Complete physics program of KLOE-2 contains: CKM unitarity and lepton universality tests, $\gamma\gamma$ physics, search for quantum decoherence and testing CPT conservation, low-energy QCD, rare kaon decays, physics of $\eta$ and $\eta^\prime$, structure of low-mass scalars, contribution of vacuum polarization to $(g-2)_{\mu}$, possible search for WIMP dark matter.
Full description of this physics program is given in ref.~\cite{amelino1}.

\subsection{The two-photon physics}
\begin{figure}[h]
\begin{tabular}{ccc}
\includegraphics[height=4cm,width=4.5cm]{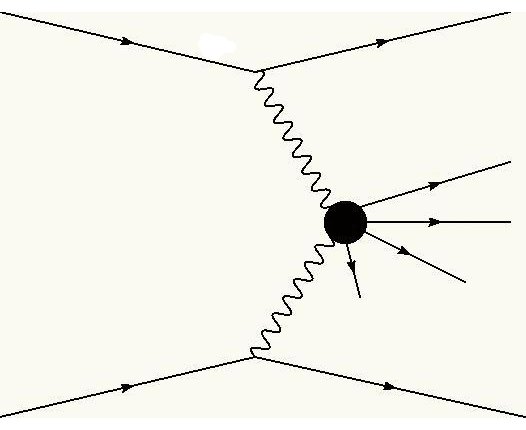} & \includegraphics[height=5cm,width=5cm]{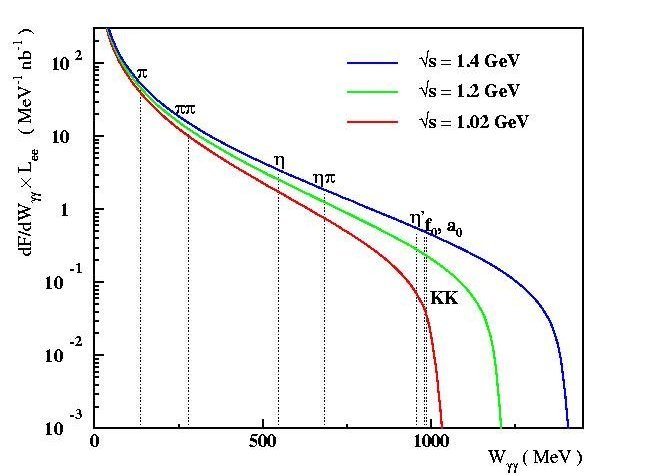} & \hspace{-5mm} \includegraphics[height=5cm,width=5cm]{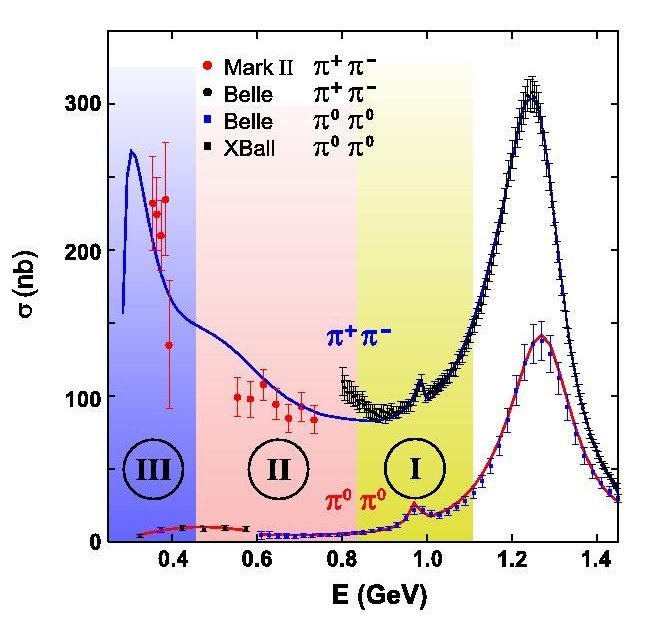}
\end{tabular}
\vspace{1mm}
\caption{Left: basic diagram describing the process $e^+e^-\rightarrow e^+e^-\gamma^\ast\gamma^\ast\rightarrow e^+e^-X$. Centre: Normalized photon flux as a function of two-photon mass and for three values of centre-of-mass energy of colliding electrons. Right: Compilation of the $\gamma\gamma\rightarrow 2\pi$ cross section \cite{pennington}.}
\label{fig3}
\end{figure}

In the reaction $e^+e^-\rightarrow e^+e^-\gamma^\ast\gamma^\ast\rightarrow e^+e^-X$, Fig.~\ref{fig3} (left), two quasi-real photons produce hadronic final state $X$ with quantum numbers $J^{PC}=0^{\pm +},2^{\pm +}$. 
Differential number of $X$ as function of the $\gamma\gamma$ mass is given as
\begin{eqnarray}
\frac{dN_X}{dW_{\gamma\gamma}}=L_{ee}\frac{dF}{dW_{\gamma\gamma}}\sigma_{\gamma\gamma\rightarrow X},
\label{eq1}
\end{eqnarray}
where the normalized flux is presented in Fig.~\ref{fig3} (centre).

Special interest is devoted to $X$ being two-meson state, either charged or neutral.
In particular, for $X=\pi\pi$ compilation of the production cross section data from Belle, Crystal Ball and Mark~II as a function of energy \cite{pennington} is displayed in Fig.~\ref{fig3} (right).
Regions marked as I, II and III, to all of which KLOE-2 can contribute, correspond to different classes of physical phenomena.
In region I, 850-1100 MeV, accurate measurements of $2\pi$ differential cross sections and any precise measurements of the cross section for $K\bar K$ and $\pi\eta$ final states are needed.
For region II, 450-850 MeV, precision measurement of $2\pi$ final states, in both charge modes, are called for, mainly for study of $\sigma/f_0$ meson.
Energy region III, i.e. 280-450 MeV, is covered by outdated measurements and calls for signifucant improvement in data statistics. This domain is crucial for the partial-wave analysis of resonances.
The $\pi^0\pi^0$ channel is generally cleaner than $\pi^+\pi^-$, being devoid of the $\mu^+\mu^-$ background, but is still contaminated by the background from $\Phi\rightarrow K^0\bar K^0$ with one escaping kaon. 
This background will be supressed by using taggers.

For $W_{\gamma\gamma}> 1$ GeV, photons reveal internal quark structure od the $f_2(1270)$ tensor meson, decaying into two pions with branching fraction 85\%. 
Therefore $\sigma_{\pi^+\pi^-}\rightarrow\sigma_{\pi^0\pi^0}$ for high energy.
For energy below 1 GeV, photons couple mainly to the two-pion final state from decay of the scalar $\sigma/f_0(600)$ and therefore $\sigma_{\pi^0\pi^0}$ becomes small.

Interesting possibility is opened by measurement of the final-state pseudoscalar mesons like $\pi$, $\eta$ and $\eta^\prime$. 
In the narrow-state approximation, the production cross section for pseudoscalar $P$ is related to its decay width into two photons $\Gamma(P\rightarrow\gamma\gamma)$.
For $P=\eta,\eta^\prime$, this width can be used for determination of the $\eta-\eta^\prime$ mixing angle and the gluonic contents of $\eta^\prime$ \cite{marsiske}, transision form factors and contribution to the light-by-light scattering in $(g-2)_\mu$ \cite{amelino1}.

\begin{figure}[t]
\begin{tabular}{cc}
\hspace{-5mm} \includegraphics[scale=.27]{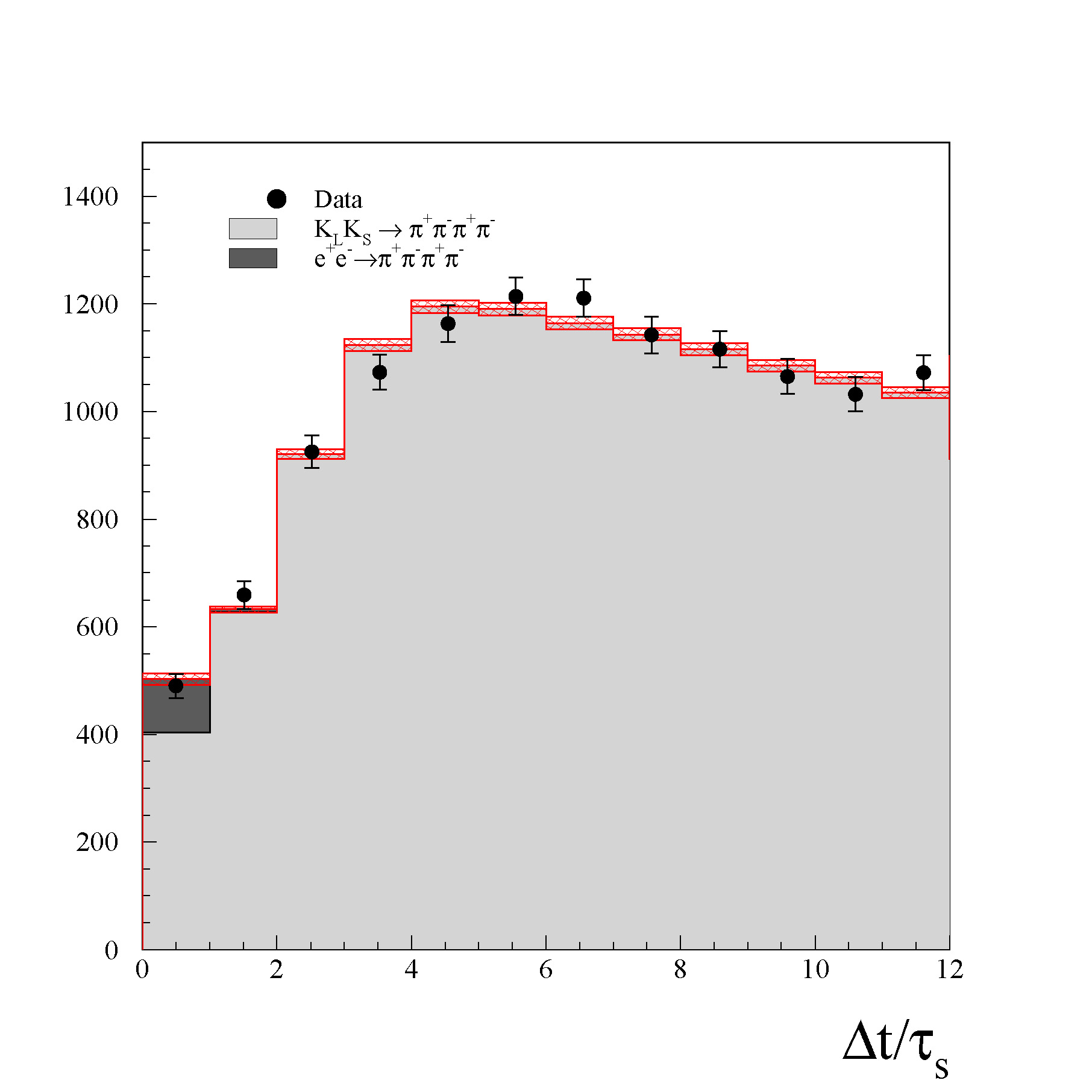} & \hspace{-8mm} \vspace{-2mm} \includegraphics[scale=.25]{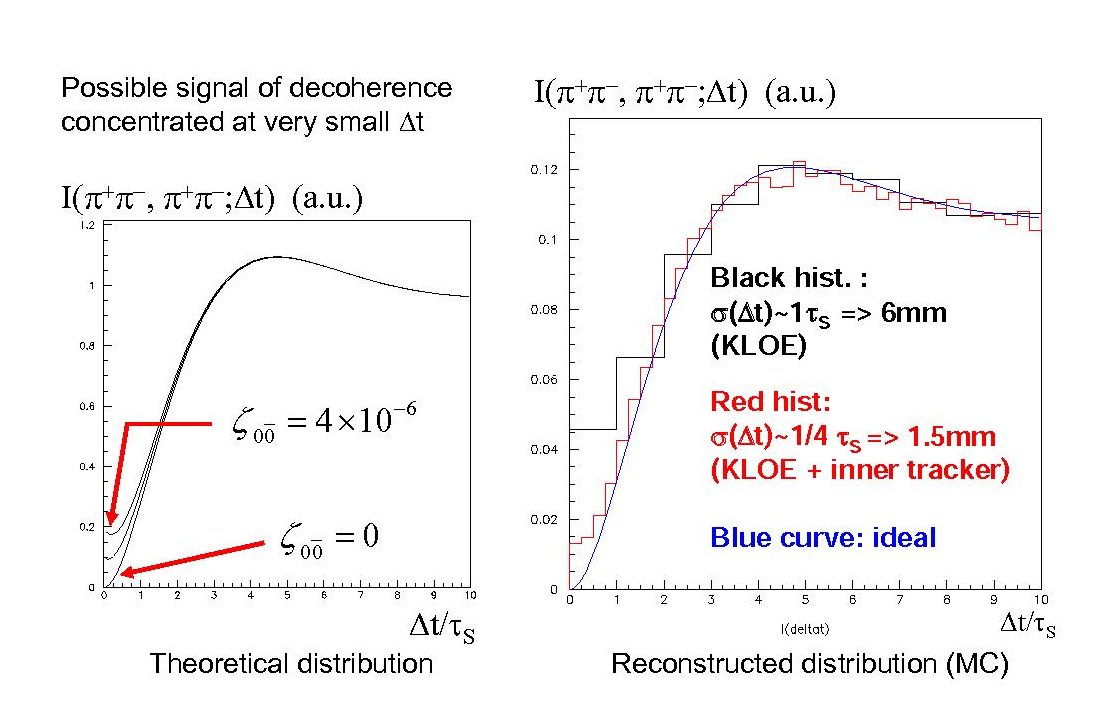}
\end{tabular}

\caption{Spectra of difference of decay times (3.3), together with backgrounds and fits, for KLOE \cite{kloe_dec} (left) and foreseen improvement of precision due to installation of inner tracker in KLOE-2 (centre and right).}
\label{fig4}
\end{figure}

\subsection{Decoherence in entangled two-kaon states}

Following time evolution of $K_{L,S}$ pairs, born back-to-back in pure quantum $1^{--}$ state from $\Phi$ decay
\begin{eqnarray}
|i\rangle=\frac{1}{\sqrt{2}}(|K_S(\vec p)\rangle|K_L(-\vec p)\rangle-|K_L(\vec p)\rangle|K_S(-\vec p)\rangle)
\label{eq2}
\end{eqnarray}
provides with a unique opportunity to test the validity of quantum mechanics \cite{didomenico}.
In the most natural and simplest approach, one investigates the time difference $\Delta t$ spectrum for kaons decaying into $f_{1,2}$ final states
\begin{eqnarray}
I(f_1,f_2;\Delta t)\sim |\eta_1|^2e^{-\Gamma_L\Delta t}+|\eta_2|^2e^{-\Gamma_S\Delta t}-2(1-\zeta)|\eta_1||\eta_2|e^{-(\Gamma_L+\Gamma_S)\Delta t/2}\cos(\Delta m\Delta t+\Delta\phi_K)
\label{eq3}
\end{eqnarray}
where $0\le\zeta\le 1$ parametrizes the interference term. 
In particular, $\zeta=1$ corresponds to the Furry hypothesis \cite{furry} of complete coherence lost. 

Hypotheses with more physics insight were subsequently proposed: dissipative decoherence with a dissipative term added to the Liouville equation for evolution of the kaon density matrix parametrized with three parameters $\alpha$, $\beta$ and $\gamma$ \cite{ellis}, and the symmetric state admixture $\omega |i^\prime\rangle=\frac{\omega}{\sqrt{2}}(|K_S(\vec p)\rangle|K_S(-\vec p)\rangle-|K_L(\vec p)\rangle|K_L(-\vec p)\rangle)$ to the initial state $|i\rangle$ (\ref{eq2}) (cf. ref. \cite{bernabeu}). 

Detection of kaons in KLOE(KLOE-2) is performed by tagging: either $K_L$ is tagged by the $K_S\rightarrow\pi\pi$ decay close to the interaction point, or $K_S$ is tagged by $K_L$ crashing in the electromagnetic calorimeter.

In Fig.~\ref{fig4} the spectra (\ref{eq3}) are presented, as measured by the KLOE Collaboration \cite{kloe_dec}, and improvement expected from KLOE-2 with vertex resolution sharpened by using the inner tracker.
Summary of the best fit values obtained by KLOE and errors expected from KLOE-2 for 25 fb$^{-1}$ of data \cite{kloe_dec} is given in Tab.~1.
As can be judged from the table, an order of magnitude improvemnt is expected for all parameters, except for the least sensitive $\alpha$.
\begin{table}[h]
\begin{center}
\begin{tabular}{||c|c|c||}
\hline\hline
Parameter & KLOE & KLOE-2 (25 fb$^{-1}$) \\
\hline
$\zeta_{LS}$         & $(0.3\pm 1.9)\times 10^{-2}$ & $\pm 0.2\times 10^{-2}$ \\
$\zeta_{00}$         & $(0.1\pm 1.0)\times 10^{-6}$ & $\pm 0.1\times 10^{-6}$ \\
$\Re\omega$          & $(-1.6^{+3.0}_{-2.1}\pm0.4)\times 10^{-4}$ & $\pm 3\times 10^{-5}$ \\
$\Im\omega$          & $(-1.7^{+3.3}_{-3.0}\pm1.2)\times 10^{-4}$ & $\pm 4\times 10^{-5}$ \\
$\alpha$             & $(-0.5\pm 2.8)\times 10^{-17}$ GeV & $\pm 2\times 10^{-17}$ GeV \\
$\beta$              & $(2.5\pm 2.3)\times 10^{-19}$ GeV & $0.2\times 10^{-19}$ GeV \\
$\gamma$             & $(1.1\pm 2.5)\times 10^{-21}$ GeV & $\pm 0.3\times 10^{-21}$ GeV \\
$\gamma$ (pos. hyp.) & $(0.7\pm 1.2)\times 10^{-21}$ GeV & $\pm 0.2\times 10^{-21}$ GeV \\
\hline\hline
\end{tabular}
\caption{The best fit values obtained by KLOE \cite{kloe_dec} for decoherence parameters, using different models of decoherence \cite{furry,ellis,bernabeu}, and errors foreseen for KLOE-2 using 25 fb$^{-1}$ of data. The last line refers to $\gamma$ with the positivity hypothesis \cite{huet}.}
\end{center}
\end{table}

\section{Acknowledgements}
This work was partially supported by the Polish MNiSW Grant No. 0469/-B/H03/2009/37, FP7 Research Infrastructure Hadron Physics 2 (INFRA 2008 227431) and FP6 Marie Curie Research and Training Network FLAVIAnet (MRTN CT 2006 035482).

\end{document}